\documentclass[preprint,aps,amsmath,byrevtex,prd,titlepage,
nofootinbib]{revtex4-2}

\usepackage{dcolumn}
\usepackage{amsmath}
\usepackage{amssymb}
\usepackage{bm}
\usepackage{hyperref}
\usepackage{graphicx}
\usepackage{slashed}
\usepackage{subfigure}

\newcommand{\beq}{\begin{equation}}
\newcommand{\eeq}{\end{equation}}
\newcommand{\eq}[1]{Eq.~(\ref{#1})}

\begin{document}

\title{Two-Loop  Electron Factor Contribution to Lamb Shift in Muonium and Positronium}

\author {Michael I. Eides}
\email[Email address: ]{meides@g.uky.edu}
\affiliation{Department of Physics and Astronomy,
University of Kentucky, Lexington, KY 40506, USA}
\author{Valery A. Shelyuto}
\email[Email address: ]{shelyuto@vniim.ru}
\affiliation{D. I.  Mendeleyev Institute for Metrology,
St.Petersburg 190005, Russia}

\begin{abstract}
We calculate hard spin-independent contributions to energy levels in muonium and positronium which are due to radiatively corrected  electron factor insertion in two-photon exchange diagrams. Calculation of these corrections is motivated by the new round of precise measurements of spin-independent transition frequencies in muonium and positronium.
\end{abstract}

\maketitle


For many years experimental and theoretical research on energy levels in muonium and positronium concentrated on hyperfine structure, see reviews in \cite{Eides:2000xc,Eides:2007exa,Karshenboim:2005iy,Tiesinga:2021myr,Adkins:2022omi}.  Now a new generation of experiments on measuring spin-independent transitions ($1S-2S$, $2S-2P$, etc.) in muonium and positronium (see, e.g., \cite{Crivelli:2018vfe,Ohayon:2021dec,yksu2018,Mu-MASS:2021uou,Crivelli:2016fjw,
Mills:2016,Cassidy:2018tgq,Gurung:2020hms,Cortinovis:2023zqi,Janka:2021xxr,Janka:2022pis,
Sheldon:2023eic,Adkins:2022omi}) is either going on or planned. Inspired by these new developments  we recently started a program of calculating hard three-loop spin-independent corrections to energy levels in muonium and positronium \cite{Eides:2021wuv,Eides:2022nda}.

\begin{figure}[h!]
\centering{
\subfigure[]
{\includegraphics[width=5cm]{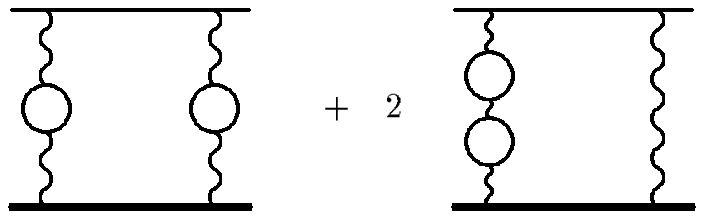}}
\hskip2cm
\subfigure[]
{\includegraphics[width=5cm]{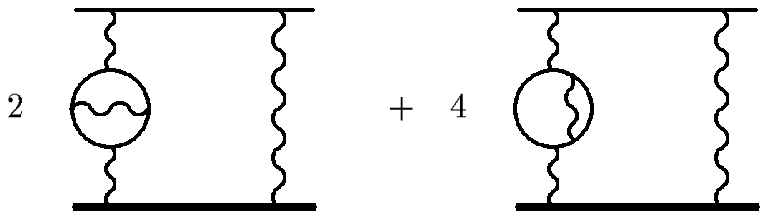}}}\\
\centering{\subfigure[]
{\includegraphics[width=12cm]{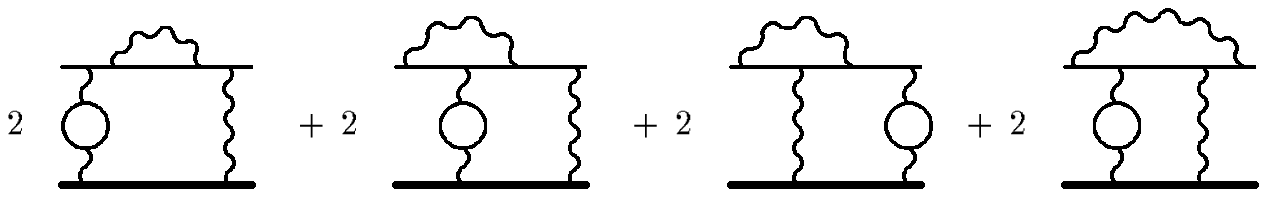}}}
\caption{
}
\label{olddiarg}
\end{figure}

There are numerous  gauge invariant sets of diagrams generating such corrections. We have already calculated contributions of three gauge invariant sets of diagrams in Fig.~\ref{olddiarg} \cite{Eides:2021wuv,Eides:2022nda}. 

\begin{figure}[h!]
\begin{center}
\includegraphics[width=10cm]{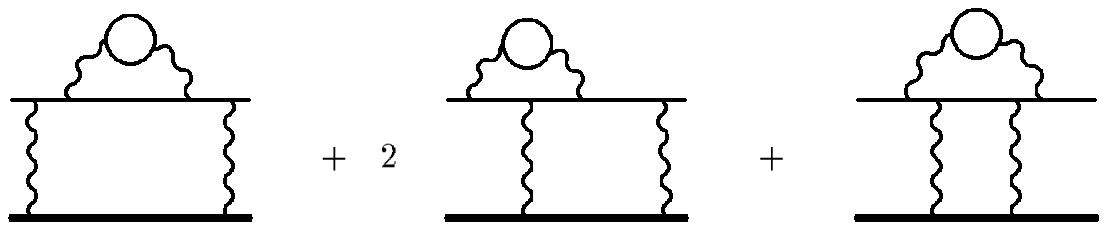}
\end{center}
\caption{
}
\label{polinsert}
\end{figure}

Below we calculate contributions to the Lamb shift in muonium and positronium generated by one more set of diagrams in Fig.~\ref{polinsert} (plus diagrams with crossed exchanged photons, which are not shown explicitly). In the case of muonium nonrecoil contribution of these diagrams was calculated long time ago \cite{Eides:1992gr}, so here we calculate only the radiative-recoil contribution.  

For any system of two electromagnetically interacting leptons with unequal (or equal) masses the hard spin-independent energy shift to the bound state energy level generated by the diagrams with two-photon exchanges is described by the integral \cite{Eides:2000kj}

\beq  \label{general}
\Delta E=-\frac{(Z\alpha)^5}{\pi n^3}m_r^3
\int {\frac{d^4 k}{i\pi^2 k^4}} \frac{1}{4} Tr \Bigl[(1 + \gamma_0 )L_{\mu \nu} \Bigr]
\frac{1}{4} Tr \Bigl[(1 + \gamma_0 )H_{\mu \nu} \Bigr]\delta_{l0},
\eeq

\noindent
where $m$ and $M$ are the masses of the leptons, $L_{\mu \nu}$ and $H_{\mu \nu}$ are the light and heavy fermion factors, respectively, $m_r=mM/(m+M)$ is the reduced mass, $Z=1$ is the charge of the heavy fermion in terms of the positron charge,  $n$ and $l$ are the principal quantum number and the orbital momentum, respectively. The  expression in \eq{general} is exact in the mass ratio, and is valid also in the case of $m=M$ (positronium).

\begin{figure}[h!]
\begin{center}
\includegraphics[width=10cm]{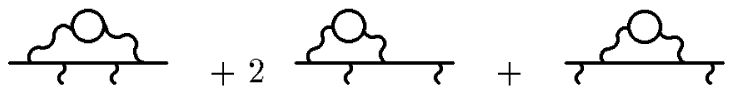}
\end{center}
\caption{
}
\label{elfact}
\end{figure}

The radiatively corrected electron factor is a sum of three terms 

\beq
L_{\mu \nu}= L_{\mu \nu}^{\Sigma} + 2L_{\mu \nu}^{\Lambda}
+ L_{\mu \nu}^{\Xi},
\eeq

\noindent
arising from the two-loop self-energy, vertex and spanning photon insertions in electron line and 
corresponding to the diagrams in Fig.~\ref{elfact}.    Respectively, the first trace in \eq{general} can be written as 

\beq
\frac{1}{4} Tr \Bigl[(1 + \gamma_0 )L_{\mu \nu} \Bigr]\equiv\frac{\alpha^2}{\pi^2 m}{\cal L}_{\mu \nu}\left(\frac{k}{m}\right)=\frac{\alpha^2}{\pi^2 m}\left[{\cal L}_{\mu \nu}^{\Sigma}\left(\frac{k}{m}\right) + 2{\cal L}_{\mu \nu}^{\Lambda}\left(\frac{k}{m}\right)
+ {\cal L}_{\mu \nu}^{\Xi}\left(\frac{k}{m}\right)\right]. 
\eeq

A photon line with the insertion of a one-loop polarization operator in Fig.~\ref{elfact}  has a natural interpretation as a massive photon propagator with the mass squared
$\lambda^2=4m^2/(1 -v^2)$. The diagrams with the polarization insertions are obtained from this massive propagator integrating over $v$ with the weight $(\alpha/\pi)v^2(1-v^2/3)/(1-v^2)$. All entries in the two-loop fermion factor except the two-loop anomalous magnetic moment and two-loop slope of the electric form factor decrease
at least as $k^2$ at $k^2\to0$. As a result, the slowly decreasing terms with the anomalous magnetic moment and slope of the electric form factor produce infrared-divergent contributions in the integral in \eq{general} for the diagrams in Fig.~\ref{polinsert}. This linear infrared divergence indicates existence of a contribution to the Lamb shift of the previous order in $Z\alpha$ that is already well known. To get rid of this spurious divergence, we subtract the terms with the two-loop anomalous magnetic moment and slope of the electric form factor from the two-loop electron factor.

The heavy line factor in \eq{general} has the form

\beq
H_{\mu \nu}=\gamma_{\mu} \frac{\hat{P} + \hat{k} + M}
{k^2 + 2Mk_0 + i0} \gamma_{\nu}+
\gamma_{\nu}  \frac{\hat{P} - \hat{k} + M}{k^2 - 2Mk_0 +
i0}\gamma_{\mu},
\eeq

\noindent
where $P=(M,{\bf 0})$ is the momentum of the particle with mass $M$.

\noindent
In the case of $m\ll M$ the heavy trace reduces  to

\beq \label{LambRR-6}
\begin{split}
\frac{1}{4} Tr \Bigl[(1 + \gamma_0 )H_{\mu \nu} \Bigr] &
\to
- \frac{1}{M}  \biggl[k^2  g_{\mu 0} g_{\nu 0}
\wp \Bigl(\frac{1}{k_0^2}\Bigr)
- \bigl(g_{\mu 0} k_{\nu} + g_{\nu 0} k_{\mu} \bigr)
\frac{1}{k_0}+ g_{\mu \nu}\biggr]\\
&
\equiv - \frac{1}{M}{\cal H}_{\mu\nu}(k)_{rec},
\end{split}
\eeq

\noindent
where ${\cal H}_{\mu\nu}(k)_{rec}$ is a dimensionless function, and $\wp \Bigl(\frac{1}{k_0^2}\Bigr)$ is the principal value integral, see \cite{Eides:2000kj,Eides:2013yxa} for the definition and properties.

The linear in mass ratio radiative-recoil contribution is obtained from \eq{general}  by the substitution in \eq{LambRR-6} \cite{Eides:2000kj}

\beq \label{ordermm}
\Delta E_{rec} =\frac{\alpha^2(Z\alpha)^5}{\pi^3 n^3}
\frac{m_r^3}{Mm}
\int {\frac{d^4 k}{i\pi^2 k^4}}{\cal L}_{\mu \nu}\left(\frac{k}{m}\right){\cal H}_{\mu\nu}(k)_{rec}.
\eeq

\noindent 
This expression will be used for calculation of the radiative-recoil contribution of the diagrams in Fig.~\ref{polinsert} in muonium.

We have shown in \cite{Eides:2022nda} that there exists a simple relationship between the integrand for the linear in mass ratio radiative-recoil corrections in \eq{ordermm}  and the integrand for the total (recoil and nonrecoil) spin-independent contribution in the case of equal masses $m=M$. Namely, it is sufficient to let $m=M$ and make the substitution ${\cal H}_{\mu\nu}(k)_{rec}\to{\cal H}_{\mu\nu}(k)_{tot}$, where 

\beq \label{eqmassintrg}
{\cal H}_{\mu\nu}(k)_{tot}={\cal H}_{\mu\nu}(k)_{rec}\frac{k_0^2}{k_0^2 -\frac{ k^4}{4m^2}}
=\frac{k^2 g_{\mu 0} g_{\nu 0}
- (g_{\mu 0} k_{\nu} + g_{\nu 0} k_{\mu})
{k_0} + g_{\mu \nu}k_0^2 }
{k_0^2 -\frac{ k^4}{4m^2}}.
\eeq

\noindent
Then the total contribution to the spin-independent energy shift of order $\alpha^7m$ generated by the diagrams in Fig.~\ref{polinsert} in the case of equal masses is given by the integral 

\beq \label{eqmss}
\Delta E_{tot} =2\frac{\alpha^2(Z\alpha)^5}{\pi^3 n^3}
\frac{m_r^3}{Mm}
\int {\frac{d^4 k}{i\pi^2 k^4}}{\cal L}_{\mu \nu}\left(\frac{k}{m}\right){\cal H}_{\mu\nu}(k)_{tot},
\eeq

\noindent
where an extra factor $2$ reflects the possibility to make  radiative insertions in both fermion lines. 

We calculated the energy shifts in the Feynman gauge for the radiative photons. The linear infrared divergences, which, as was explained above, indicate the presence  of the contributions of the previous order in $Z\alpha$, were omitted, and the spurious logarithmic infrared divergences cancelled in the sum of diagrams in Fig.~\ref{polinsert}. Using \eq{ordermm} we obtain radiative-recoil correction in muonium 

\beq \label{VP-eLine-01}
\Delta E^{(Mu)}=\left(J^{(Mu)}_{\Sigma} + 2J^{(Mu)}_{\Lambda} + J^{(Mu)}_{\Xi }\right)\frac{\alpha^2(Z\alpha)^5}{\pi^3n^3}\frac{m}{M}
\left(\frac{m_r}{m}\right)^3m\delta_{l0}.
\eeq

\noindent
Calculations are similar to the ones in  \cite{Eides:2021wuv,Eides:2022nda}, and  the infrared finite contributions of the diagrams in Fig.~\ref{polinsert} are as follows

\beq
J^{(Mu)}_{\Sigma}=0.10602(3),\quad 2J^{(Mu)}_{\Lambda }=-0.07644(2),\quad J^{(Mu)}_{\Xi}=0.07373(3).
\eeq

\noindent
The total radiative-recoil contribution to the Lamb shift in muonium from the diagrams in Fig.~\ref{polinsert} is

\beq \label{elfVP}
\Delta E^{(Mu)}=0.10332(3)\frac{\alpha^2(Z\alpha)^5}{\pi^3n^3}\frac{m}{M}
\left(\frac{m_r}{m}\right)^3m\delta_{l0}.
\eeq

To calculate the spin-independent contribution to the energy shift in positronium we use the expression in \eq{eqmss} 

\beq \label{ps}
\Delta E^{(Ps)}=\left(J^{(Ps)}_{\Sigma} + 2J^{(Ps)}_{\Lambda} + J^{(Ps)}_{\Xi }\right)\frac{\alpha^7}{\pi^3n^3}\frac{m}{4}\delta_{l0},
\eeq

\noindent 
where the infrared finite contributions of separate diagrams in the Feynman gauge are

\beq \label{poscpnvem}
J^{(Ps)}_{\Sigma}=-0.11001(2),\quad 2J^{(Ps)}_{\Lambda}=-0.07969(2),\quad J^{(Ps)}_{\Xi}=-0.32816(1).
\eeq

\noindent
Finally, the contribution to the Lamb shift in positronium from the diagrams in Fig.~\ref{polinsert} is

\beq 
\Delta E^{(Ps)}=-0.12947(3)\frac{\alpha^7m}{\pi^3n^3}\delta_{l0}.
\eeq

Combining the results in \eq{elfVP} with our earlier results for muonium  \cite{Eides:2021wuv} we obtain the total radiative-recoil contribution  of the diagrams in Fig.~\ref{olddiarg} and Fig.~\ref{polinsert}

\beq \label{newresultmu}
\Delta E^{(Mu)}=-11.3275(2)\frac{\alpha^2(Z\alpha)^5}{\pi^3 n^3}\frac{m}{M}\left(\frac{m_r}{m}\right)^3m\delta_{l0}.
\eeq

\noindent
The contribution to the Lamb shift in positronium  generated by the diagrams in Fig.~\ref{olddiarg} \cite{Eides:2021wuv,Eides:2022nda} and Fig.~\ref{polinsert} is 

\beq \label{newresultpos}
\Delta E^{(Ps)}=0.8057(2)
\frac{\alpha^7m}{\pi^3n^3}\delta_{l0}.
\eeq

\noindent
The contributions in \eq{newresultmu} and \eq{newresultpos} are too small to play a significant role for the results of the ongoing experiments, they are at the level of a few tenths of kHtz and a few kHz, respectively. However, we expect that these corrections will become phenomenologically relevant in the future with further improvements of the experimental accuracy.

There are other gauge-invariant sets of three-loop diagrams which arise as radiative corrections to the two-photon exchange diagrams, see, e.g.,  \cite{Eides:2013yxa}.  Hard spin-dependent  corrections generated by these diagrams are already calculated, see, e.g., the review in \cite{Eides:2016qhf} and references therein. Respective spin-independent corrections remain at this time unknown, and we hope to calculate them in the near future.

\acknowledgments

Work of M. I. Eides was supported by the NSF grant PHY- 2011161.

\end{document}